\documentclass[twocolumn]{aastex63}
\usepackage{amsmath}

\received{June 1, 2019}
\revised{January 10, 2019}
\accepted{\today}
\submitjournal{ApJ}

\shorttitle{D1 census}
\shortauthors{Shi et al.}
\begin{document}

\title{A detailed study of massive galaxies in a protocluster at z=3.13 }

\correspondingauthor{Ke Shi}
\email{keshi@xmu.edu.cn}

\author{Ke Shi}

\affiliation{Department of Astronomy, Xiamen University, Xiamen, Fujian 361005, China}

\author{Jun Toshikawa}

\affiliation{Institute for Cosmic Ray Research, The University of Tokyo, Kashiwa, Chiba 277-8582, Japan}

\affiliation{Department of Physics, University of Bath, Claverton Down, Bath, BA2 7AY, UK}

\author{Zheng Cai}

\affiliation{Department of Astronomy, Tsinghua University, Beijing 100084, China}

\author{Kyoung-Soo Lee}

\affiliation{Department of Physics and Astronomy, Purdue University, 525 Northwestern Avenue, West Lafayette, IN 47907, USA}

\author{Taotao Fang}
\affiliation{Department of Astronomy, Xiamen University, Xiamen, Fujian 361005, China}

\begin{abstract}
We present a detailed study of Near-IR selected galaxies in a protocluster field at $z=3.13$. Protocluster galaxies are selected using the available mutliwavelength data with the photometric redshifts (photo-$z$) at $2.9<z<3.3$, reaching a mass completeness of $\simeq10^{10}~\mathrm{M_\sun}$. Diverse types of galaxies have been found in the field including normal star-forming galaxies, quiescent galaxies and dusty star-forming galaxies. The photo-$z$ galaxies form two large overdense structures in the field, largely overlapping with the previously identified galaxy overdensities traced by Ly$\alpha$ emitters (LAEs) and Lyman break galaxies (LBGs) respectively. The northern overdensity consists of a large fraction of old and/or dusty galaxy populations, while the southern one is mainly composed of normal star-forming galaxies which are spatially correlated with the LAEs. This agrees with our previous study arguing the spatial offset of different galaxy overdensities may be due to halo assembly bias. Given the large end-to-end sizes of the two overdensities, one possibility is that they will form into a supercluster by the present day. We also find strong evidence that the star-formation activities of the galaxies in the overdense protocluster regions are enhanced in comparison to their field counterparts, which suggests an accelerated mass assembly in this protocluster.

\end{abstract}



\section{Introduction} \label{sec:intro}
Hierarchical structure formation theory predicts that structures form in a bottom-up way, such that initial small density fluctuations give rise to proto-stars which form into first galaxies. These galaxies subsequently grow larger and become more massive via mergers and accretion, followed by the formation of groups, clusters and superclusters of galaxies \citep{White78}. As the densest large-scale structures of the universe, galaxy clusters provide us with unique laboratories to study how galaxy formation proceeds in dense environments. 

It is well known that galaxy formation is strongly affected by the local environments in which galaxies reside. In the local universe, cluster galaxies form a tight `red sequence' \citep{Visvanathan77,Bower92,Stott09} and obey the `morphology-density' relation \citep{Dressler80,Dressler97, Goto03, Kauffmann04}, in a sense that cluster galaxies are typically red massive ellipticals while young star-forming galaxies such as spiral galaxies tend to reside in the field. Furthermore, observational evidence suggest that cluster galaxies experienced an accelerated mass assembly followed by a swift shutdown of their star formation, and evolve passively till the present day \citep[e.g.,][]{Stanford98,Thomas05,Snyder12,Mart18}. 

The dominant population of massive quiescent galaxies in clusters in the local universe also implies that the star formation-density relation may be reversed at higher redshift ($z>1$). Indeed, studies of distant clusters and progenitors of clusters (`protoclusters') have shown that star formation activities in dense environments are enhanced relative to the field \citep[e.g.,][]{Elbaz07,Cooper08,Tran10,Koyama13,Alberts14,Shimakawa18}. However, exactly when this reversal occurs and a detailed assembly history of cluster galaxies are still largely unknown \citep[e.g.,][]{Snyder12,Lemaux18}. In order to better understand the formation and subsequent quenching of cluster galaxies, we need to directly witness protoclusters and their galaxy constituents at high redshift ($z>2$), the epoch when the cosmic star formation activity is about to reach its peak \citep{Madau14}. 

Distant protoclusters are rare, the largest ones (those which will evolve into a Coma-size cluster of mass $\gtrsim 10^{15} \mathrm{M_\sun}$) have a comoving space density of only $\sim 2 \times 10^{-7}$ Mpc$^{-3}$ \citep{Chiang13}. They are not virialized yet and usually span large angular sizes of 10$\arcmin$-30$\arcmin$ in the sky \citep{Chiang13,Muldrew15}, which makes it observationaly difficult and expensive to conduct a systematic search. So far, only several tens of protoclusters have been confirmed \citep[e.g.,][]{Overzier16,Harikane19}. 

Many studies have used distant radio galaxies or quasars as signposts of overdense regions and identified an overdensity of emission line galaxies such as Ly$\alpha$ emitters (LAEs) or H$\alpha$ emitters (HAEs) near the radio galaxies or quasars, followed up by spectroscopy to confirm these protocluster candidates \citep[e.g.,][]{Pentericci00,Kurk04,Kashikawa07,Venemans07,Kuiper11,Hatch11,
Hayashi12,Wylezalek13,Cooke14,Adams15}. However, it should be noted that there are many other studies finding no association between these signposts and protoclusters \citep[e.g.,][]{Husband13,Uchiyama18,Shi191}, which imply they are biased tracers of the underlying matter distribution. Another popular way to search for protoclusters is to resort to extensive spectroscopy of `blank fields' \citep[e.g.,][]{Steidel98,Steidel05,Toshikawa12,Lemaux14, Lee14, Cucciati14, Dey16, Toshikawa16, Cucciati18, Lemaux18, Jiang18}. Many of these protoclusters are found by pre-selecting overdense regions traced by star-forming galaxies such as LAEs or Lyman break galaxies (LBGs) with followup spectroscopic confirmations. A new promising way to select and map protoclusters is using hydrogen gas absorption \citep[e.g.,][]{Cai16,Lee16,Cai17b}, which is based on the fact that distant overdense regions contain not only large concentration of galaxies but also a large quantities of cold or warm gas that can be detected via absorption against luminous background sources such as QSOs.

A critical element in understanding galaxy and cluster formation is a detailed study of protocluster constituents. Studying how different galaxy populations are distributed within the large-scale structure is necessary to understand how galaxy formation is affected by its local environment. For instance, luminous Ly$\alpha$
nebulae are often found to be located at the outskirts or  intersections of
the densest regions of a protocluster \citep[e.g.,][]{Matsuda05,Badescu17,Cai17a,Shi192}. Powerful AGNs and dusty star-forming galaxies have also been reported to reside in abundance in dense protocluster environments \citep[e.g.,][]{Ivison00,Lehmer09,Umehata15,Casey15,Hung16,Casey16,Oteo18,Kubo19}. Investigating these sources can give us invaluble hints on how cluster ellipicals and the brightest cluster galaxy (BCG) are assembled in dense environments.

In this paper, we present a multiwavelength study of galaxies in and around a protocluster in the D1 field of the Canada-France-Hawaii-Telescope Legacy Survey (CFHTLS). This protocluster, dubbed `D1UD01', was originally discovered using the surface density of LBGs at $z\sim3-5$ \citep{Toshikawa16}. Follow-up spectroscopy confirmed five galaxies at $z=3.13$ within 1 Mpc of one another, suggesting the presence of an overdense structure. In \cite{Shi192}, we conducted a narrow-band survey to search for LAEs in the D1 field, finding a significant galaxy overdensity ($\delta=3.3$) located near the spectroscopic sources, suggesting a total mass of $\approx10^{15}$ M$_\sun$ comparable to that of the Coma cluster. Interestingly, the LAE overdensity is spatially segregated from the LBG overdensity, which suggests that different types of galaxies are probably biased tracers of the underlying dark matter halos that formed at different epochs (halo assembly bias). Motivated by these findings, here we conduct a detailed census of the galaxies constituents in this field, with the purpose of unveil the spatial configuration of the protocluster as well as to study the environmental impacts on galaxy formation in this protocluster.

This paper is organized as follows. In Section~\ref{sec2} we describe the data and methods used to select the protocluster galaxies.  We study different types of galaxies in details in Section~\ref{sec3}. In Section~\ref{distribution} we measure the spatial distributions of galaxies in the field  and identify two possible overdense protocluster regions. We discuss the environmental effects on galaxy properties and examine the difference of the two overdensities in terms of their galaxy constituents in Section \ref{dis}. A search for rare sources in the protocluster regions is also presented. We summarize our results in Section~\ref{sum}. Throughout this paper we use the WMAP9 cosmology ($\Omega_M=0.29, \Omega_\Lambda=0.71, \sigma_8=0.83, h=0.69$) from \cite{Hinshaw13}. All magnitudes are given in the AB system \citep{Oke83}. Distance scales are given in comoving units unless noted otherwise.

\section{Data and Analysis} \label{sec2}

\subsection{Data and photometry} \label{data}
In this work, we make use of publicly available multiwavelength data including the deep optical $ugriz$ images from the CFHTLS Deep Servey \citep{Gwyn12} and the 
near-IR $JHK_S$ bands from WIRCam Deep Survey (WIRDS) \citep{Bielby12}. We also use the \textit{Spitzer} data from the Spitzer Wide-area InfraRed Extragalactic survey \citep[SWIRE:][]{Lonsdale03} and the \textit{Spitzer} Extragalactic Representative Volume Survey \citep[SERVS:][]{Mauduit12}. The former includes 5.8$\micron$, 8.0$\micron$ and 24$\micron$ bands while the latter taken as part of post-cryogenic IRAC observations includes 3.6$\micron$ and 4.5$\micron$ bands only, which are deeper than the SWIRE counterparts. The photometric depths of CFHTLS and WIRDS data are measured from the sky fluctuations by placing 2$\arcsec$ diameter apertures in random image positions. The depths of \textit{Spitzer} data are measured in \cite{Vaccari16}. Table \ref{table1} summarizes the data sensitivity and image quality in this paper.

We resample the \textit{Spitzer} IR data to have the same pixel scale of 0.$\arcsec$186 as the optical CFHTLS and near-IR WIRDS data. To facilitate the comparison with the LAE overdensity found in the D1 field, all the images are trimmed to have the same dimension as the narrow-band $o3$ image used in \cite{Shi192} and the identical masks, with an effective area of 0.32 deg$^2$. 

We create a multiwavelength photometric catalog as follows. First, to accurately measure the photometry, we smooth the WIRDS images to match the broader PSFs of the CFHTLS data. To do so, the radial profile of the PSF in each image is approximated by a Moffat function with the measured seeing FWHM. A noiseless convolution kernel between the low and high-resolution images is then derived using the Richardson-Lucy deconvolution algorithm \citep{Richardson:72}. Each WIRDS image data is then convolved with its respective kernel to create a smoothed image that is PSF matched with the CFHTLS data. The LAE overdensity discovered in \cite{Shi192} lies near the edge the WIRDS images: 20\% of the area has no $K_S$ band coverage and additional 10\% has only partial coverage ($<$50\% of the full exposure), which limits our comprehensive study of massive galaxies in this protocluster field. Therefore in this work we base our study on the \textit{Spitzer} IRAC 3.6$\micron$ detection. At $z=3.13$, the 3.6$\micron$ mainly samples the rest-frame optical-NIR emission which enables the measurement of the stellar masses of galaxies.

Source detection and photometric measurements in the $ugrizJHK_S$ bands are carried out by running the SExtractor software \citep{Bertin96} in dual mode on the PSF matched images with the $i$ band data as the detection band. The SExtractor parameter MAG\_AUTO is used to estimate the total magnitude, while colors are computed from fluxes within a fixed isophotal area (i.e., FLUX\_ISO). As the images are PSF matched, aperture correction in all bands is assumed to be the difference between MAG\_AUTO and MAG\_ISO measured in the detection band.

 As for the  \textit{Spitzer} images, since the PSFs of the IRAC and MIPS images are much broader ($\approx$ 2$\arcsec$  and 6$\arcsec$ respectively), source blending on these images is a severe problem. In order to obtain accurate and unbiased measurement of fluxes and colors on the \textit{Spitzer} images, we utilize the T-PHOT software \citep{Merlin15,Merlin16} which performs ``template-fitting''
photometry on the low-resolution image using the information of high-resolution image and catalog. In our case, the $i$ band image and catalog are used as the input priors of T-PHOT while the low-resolution \textit{Spitzer} images are analysized to obtain precise photometry. We notice that the resultant \textit{Spitzer} fluxes derived by T-PHOT do not strongly depend on the based priors. For a test purpose, we also do a similar analysis using $r$ band as prior, and obtain the corresponding \textit{Spitzer} photometry. We cross-match the $r$ band based catalog with that of the $i$ band, finding that the 3.6$\micron$ magnitude difference between the two has only a mean value of $\sim$0.03 with a standard deviation of 0.06. Therefore we are assured that our T-PHOT photometry is robust and unbiased.

Finally, all photometric catalogs are merged together to create a multiwavelength catalog. In this work, to secure the measurement of the stellar masses of the galaxies, we focus on the sources with 3.6$\micron$ magnitudes smaller than 23.04 (i.e., $>$ 5$\sigma$ detection limit). In the end, 31,218 sources are selected in the final catalog.

\begin{deluxetable}{cccc}[h]
\tablecaption{Data Set \label{table1}}
\tablehead{
\colhead{Band} & \colhead{Instrument} & \colhead{Limiting magnitude\tablenotemark{a}} & \colhead{FWHM} \\
\colhead{} & \colhead{} & \colhead{(5$\sigma$,AB)} & \colhead{($\arcsec$)}
}
\startdata
$u$ & MegaCam/CFHT & 27.50 & 0.80\\
$g$ & MegaCam/CFHT & 27.82 & 0.80\\
$r$ & MegaCam/CFHT & 27.61 & 0.80\\
$i$ & MegaCam/CFHT & 27.10 & 0.80\\
$z$ & MegaCam/CFHT & 26.30 & 0.80\\
$J$ & WIRCam/CFHT & 24.80 & 0.68\\
$H$ & WIRCam/CFHT & 24.50 & 0.62\\
$K_S$ & WIRCam/CFHT & 24.52 & 0.67\\
3.6 $\micron$ & IRAC/\textit{Spitzer} & 23.04 & 1.80\\
4.5 $\micron$ & IRAC/\textit{Spitzer} & 22.83 & 1.80\\
5.8 $\micron$ & IRAC/\textit{Spitzer} & 19.66 & 1.90\\
8.0 $\micron$ & IRAC/\textit{Spitzer} & 19.50 & 2.20\\
24 $\micron$ & MIPS/\textit{Spitzer} & 17.55 & 5.90\\
\enddata
\tablenotetext{a}{ 5$\sigma$ limiting magnitude measured in a 2$\arcsec$ diameter aperture for the CFHT data, while for the \textit{Spitzer} data the depths are measured in \cite{Vaccari16}. 
}
\end{deluxetable}

%To facilitate the comparison with the LAE overdensity found in the D1 field, we trimmed the broad band images to match the dimension of the narrow-band image

\subsection{Photometric Redshift and Spectral Energy Distribution Fitting} \label{sedfitting}
We derive photometric redshift for each object in the catalog via the spectral energy distribution (SED) fitting technique using the CIGALE  software \citep{Noll09, Boquien19}. Based on an energy balance principle (the energy emitted by dust in the mid- and far-IR exactly
corresponds to the energy absorbed by dust in the UV-optical
range), CIGALE builds composite stellar population models from various single stellar population models, star formation histories, dust attenuation laws, etc. The model templates are then fitted to the observed fluxes of galaxies from far-ultraviolet to the radio domain, and physical properties are estimated using a Bayesian analysis.

For the SED templates, we use the stellar population synthesis models of \cite{BC03} , \citet{Calzetti00} reddening law with E(B-V) values
ranging from 0 to 2 in steps of 0.1 mag, the solar metallicity,
and  \cite{Chabrier03} initial mass function. We use the delayed star formation history (SFR $\propto$ t $\times$ exp[-t/$\tau$]) with star-forming time scale $\tau$ ranging from 0.1 to 10 Gyr. The age of the main stellar population ranges from 100 Myr to 10 Gyr, with finer grids up to 2 Gyr, after which large grids are used in order to save computation time, as in this work we are only interested in selecting $z\sim3$ galaxies. Nebular emission is also included and dust emission is modeled by \cite{Dale14}. The input redshifts are set to be between 0.1 and 5.0 in steps of 0.1. In addition, for the 24 $\micron$ detected sources, we also include the AGN models from \cite{Fritz06} to better constrain the dust emission and AGN contribution.

We compare our photometric redshift (photo-$z$) measurements with the spectroscopic redshifts from the VIMOS VLT Deep Survey \citep[VVDS:][]{LeFvre13} and VIMOS Ultra-Deep Survey \citep[VUDS:][]{Lefvre15}. The precision of the photometric redshift is measured using the normalised median absolute deviation defined as $\sigma_z=1.48\times$ median($\mid\Delta_z\mid$/(1+$z_\mathrm{spec}$)), where $\Delta_z=z_\mathrm{spec}-z_\mathrm{phot}$. This scatter measurement corresponds to the rms of a Gaussian distribution and is not affected by catastrophic outliers (i.e., objects with $\mid\Delta_z\mid$/(1+$z_\mathrm{spec}$)$>0.15$) \citep{Ilbert06,Laigle16}. 

We cross-match our sample with the VVDS and VUDS catalog and find 3,685 sources have spectroscopic redshifts. For all these sources, we obtain $\sigma_z=0.12$. The number of catastrophic failures take up to 18\% in these sources. 
The mean photometric redshift error derived by CIGALE is $\Delta z \sim0.2$, therefore we select 532 galaxies with photo-$z$ measurements of $2.9<z_\mathrm{phot}<3.3$ as potential protocluster galaxy candidates, among which 75 have spectroscopic redshifts, yielding $\sigma_z=0.06$. The reason why $\sigma_z$ becomes smaller for these protocluster galaxy candidates is that our SED modelling is tuned to select high-$z$ galaxies as described previously. We visually inspect the 532 sources and remove those with potential contamination in the photometry, including those severely blended with nearby bright sources and near the boarders of the images. We check the locations of the removed sources, confirming they are relatively randomly distributed that we do not particularly remove the galaxies in the overdense regions due to the blending issue. We also remove possible M-dwarf stars by  inspecting their spectra. In the end, 356 galaxies are selected as our photo-$z$ protocluster galaxy candidates.

We fix the best-fit photo-$z$ of the protocluster galaxy candidates, using the spectroscopic redshift when available, and re-fit their SEDs using CIGALE with the same configuration to determine their physical properties such as stellar mass, dust corrected star formation rate (SFR) and color excess of stellar continuum E(B-V), etc. The masses of the galaxies are best determined with an average error of 0.10 dex, while the errors of SFRs are relatively larger, with an average value of 0.35 dex.

For the 356 photo-$z$ galaxies, we also estimate their stellar mass completeness using an empirical method \citep{Pozzetti10,Ilbert13,Laigle16}. For each  galaxy, we compute the lowest stellar mass $M_\mathrm{lim}$ it would need to be detected at the given IRAC magnitude limit [3.6]$_\mathrm{lim}=23.04$:
\begin{equation*}
\mathrm{log}(M_\mathrm{lim})=\mathrm{log}(M)-0.4([3.6]_\mathrm{lim}-[3.6]),
\end{equation*}
then the stellar mass completeness limit corresponds to the mass above which 90\% of the galaxies lie. The resultant mass completeness limit is $\mathrm{log}(M_\mathrm{lim})=9.9$ in our photo-$z$ sample.

Finally, our photo-$z$ galaxies lie around at $z\approx3.1$ where the $K_S$ band photometry could be potentially contaminated by the [O~{\sc iii}]$\lambda\lambda$4959,5007 nebular emission lines, which could possibly affect the stellar mass measurement. For example, \cite{Schenker13} measured the rest-frame
[O~{\sc iii}] equivalent widths (EWs) for a sample of $3.0<z<3.8$ LBGs and determined an average value of 250 \AA. At $z\approx3.1$, this leads to an overestimate of $K_S$ band continuum flux density by 0.3 magnitude. However, they also noticed that if use SED-fitting to derive the physical properties, there is no significant change in the stellar mass when the [O~{\sc iii}] emission is corrected (stellar mass is only reduced by 3\%). This is because the IRAC 3.6$\micron$ and 4.5$\micron$ data provide important information of the stellar component redward of the Balmer break and are not contaminated by nebular emissions at $3<z<4$. Since our photo-$z$ galaxies all have secure 3.6$\micron$ detection, the contamination from [O~{\sc iii}] is expected to be less severe. In addition, \cite{Malkan17} noticed there is an anti-correlation between the stellar mass and [O~{\sc iii}] EW for LBGs at $z\sim3$: the higher the mass, the smaller the EW. According to their relation, our mass-selected sample at $M_{\star}>10^{9.9} M_\sun$ have a typical EW of $\sim100$ \AA, corresponding to a flux contamination of 0.13 mag. Thus the influence of nebular emission on the derived physical properties such as stellar mass would be minimal, considering the robust 3.6$\micron$ detection and high mass galaxies our sample have.

\section{Diverse galaxy populations in the protocluster field} \label{sec3}

\subsection{Selection of different galaxy populations} \label{sec:selection}

One of the main focus of this paper is to study the diverse galaxy populations in this protocluster field, in order to better understand the environmental impacts on galaxy evolution. To do so, we classify our photo-$z$ galaxies using  a $J-K_S$ versus $[3.6]-[4.5]$ color-color diagram, which is similar to those in the literature \citep[e.g.,][]{Labbe05,Papovich06,Nayyeri14,Shi191,Girelli19}. 

\begin{figure*}[ht!]
\epsscale{1.1}
\plotone{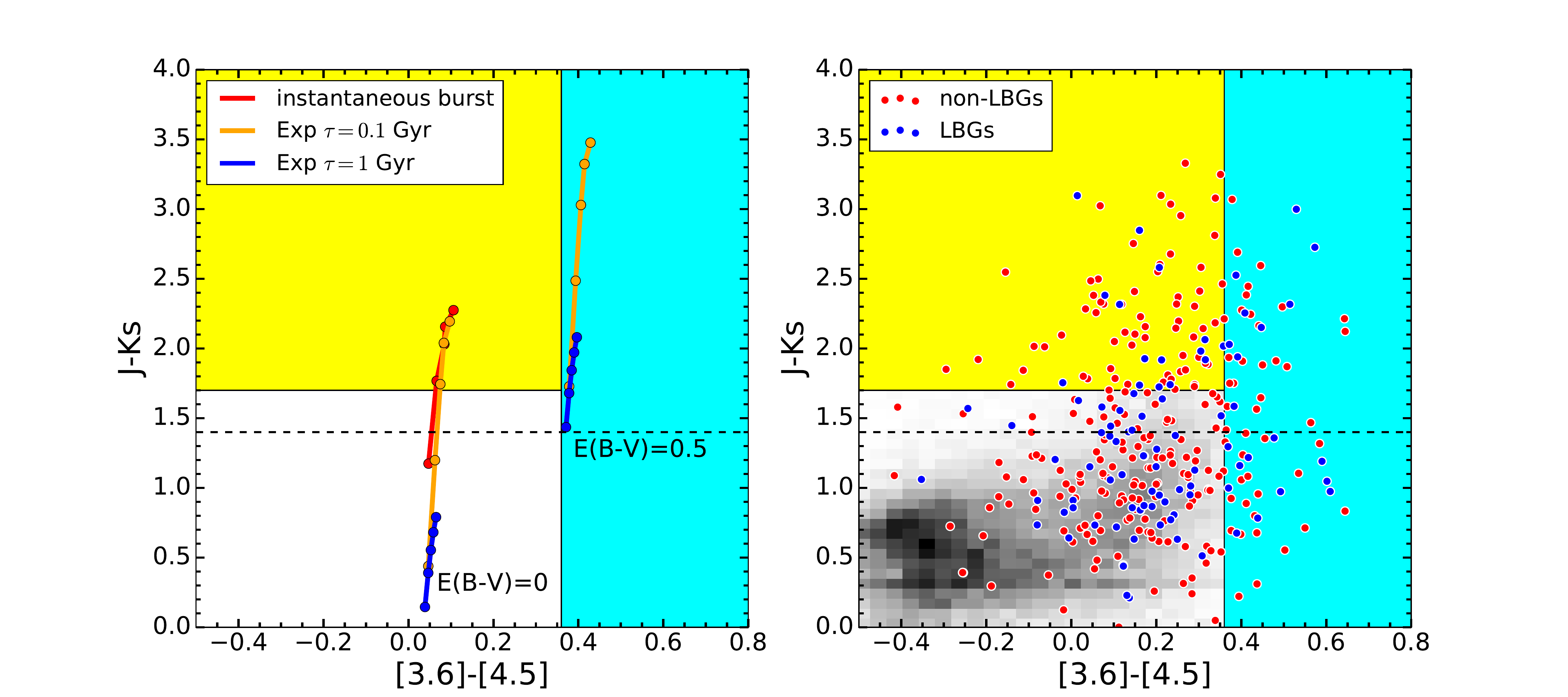}
\caption{
 {\it Left:} The color evolution of different SFH models at $z=3.13$ is shown for two dust reddening parameters E(B-V)=0 and 0.5 in the $J-K_S$ vs [3.6]$-$[4.5] color-color diagram.  The circles in each model track mark the population age of 0.2 to 1.0 Gyr in step of 0.2, from bottom to top. The yellow region represents our selection criteria for the quiescent galaxy candidates. The cyan area marks our selection of dusty star-forming galaxies while the white region represents that for normal star-forming galaxies. The dashed line marks the selection criterion for DRGs.
 {\it Right:} The photo-$z$ selected galaxies are shown in the diagram. The blue circles are those satisfying the LBG selection criteria while the red circles are non-LBGs. The gray shades show the distribution of all 3.6$\micron$ detected sources.
}
\label{fig:color}
\end{figure*}

The left panel of Figure~\ref{fig:color} shows our selection of different galaxy populations using the two-color diagram. Utilizing the EZGAL software \citep{Mancone12} with the stellar population
synthesis models of \cite{BC03} and \cite{Chabrier03} initial mass function, we compute the theoretical models of different star formation histories (SFHs) and dust reddening. Three SFHs are considered: (1) an instantaneous burst; (2) exponentially declined model (SFR $\propto$ exp[- t/$\tau$] with $\tau=0.1$ Gry; (3) exponentially declined model with $\tau=1.0$ Gyr. The ages of galaxies at the protocluster redshift $z=3.13$ are also indicated in the color tracks.

 Based on Figure~\ref{fig:color}, we classify galaxies with $J-K_S>1.7$ and $[3.6]-[4.5]<0.36$ as quiescent galaxies. The red $J-K_S$ color imposes that a strong Balmer/4000 $\mathrm{\AA}$
break fall between the $J$ and $K_S$ bands at the protocluster redshift. As can be seen in the figure, this criterion tends to select galaxies of relatively old ages ($>0.4$ Gyrs for instantaneous burst SFH model and $>0.6$ Gyrs for declined SFH model of $\tau=0.1$) with the absence of dust. Meanwhile, the $[3.6]-[4.5]$ color requires that the rest frame optical-NIR continuum slope at
$\lambda=8000-10000$ $\mathrm{\AA}$ be relatively flat, ensuring that the red
$J-K_S$ color is not due to dust reddening. It is noted that our selection of quiescent galaxies also fully incorporates the distant red galaxies (DRGs) criterion ($J-K_S>1.4$) at $2<z<4$ \citep{Franx03,van03}. Moreover, \cite{Girelli19} recently also used the same colors to select quiescent galaxies at $2<z<4$ with very similar criteria as ours. Using our criteria, 81 galaxies are selected as quiescent galaxy candidates in our sample.

In addition to passive galaxies, objects with $[3.6]-[4.5]>0.36$ are classified as dusty star-forming galaxies, because majority of these galaxies have colors consistent with continuous SFH with high dust reddening E(B-V)$\geqslant$0.5. Normal star-forming galaxies are selected to be in the region of $J-K_S<1.7$ and $[3.6]-[4.5]<0.36$, as they are mostly consistent with continuous SFH models with mild dust obscuration. In the end, 65 galaxies are classified as dusty star-forming and 210 are selected as normal star-forming galaxies in the sample.

In the right panel of Figure~\ref{fig:color}, we plot all our photo-$z$ galaxies in the color-color diagram. We also divide the sample into two catagories: LBGs which satisfy the drop-out selection criteria used in \cite{Shi192}; non-LBGs that do not satisfy the LBG criteria. Among the 356 photo-$z$ galaxies, 116 are LBGs which account for 33\% of the entire sample. The majority of the LBGs (77, account for 66\%) lie in the region of normal star-forming galaxies while 21 are distributed in the region of dusty star-forming galaxies and only 18 are classified as quiescent galaxy candidates. Our results agree with the general expectation that LBGs are usually star-forming galaxies with little or moderate dust obscuration \citep[e.g.,][]{Giavalisco02}.

We also cross-match our photo-$z$ galaxies with the LAEs in \cite{Shi192}, finding three counterparts. These galaxies are all UV-bright sources that have luminosities log(L$\mathrm{_{UV}}$) $>$ 28.5 erg s$^{-1}$ Hz$^{-1}$ at the rest-frame 1700 \AA. In comparison, the entire LAE sample has an average UV luminosity of $\sim28.0$ erg s$^{-1}$ Hz$^{-1}$. Therefore these galaxies are among the most UV-luminous LAEs in the field which are also detected in 3.6 $\micron$. In particular, one galaxy is located in the LAE overdensity found in \cite{Shi192}. SED-fitting suggests it has a stellar mass of $10^{10.5} ~\mathrm{M_\sun}$ with SFR of only 1$~\mathrm{M_\sun}$ yr$^{-1}$ and belongs to the quiescent galaxy population. Having such a high mass and low SFR, this LAE may be a rare one and worth further investigating in future observation. The remaining LAEs are not detected in 3.6$\micron$ and therefore not selected in the photo-$z$ catalog.

To sum up,  our photo-$z$ sample includes a large fraction of massive galaxies that have been missed from the rest-frame UV selected star-forming galaxies such as LBGs and LAEs. This highlights the importance of using rest-frame optical-NIR selection to study the high-mass end of the stellar mass function.

\subsection{Physical properties of different galaxy populations}
\begin{figure*}[ht!]
\epsscale{1.2}
\plotone{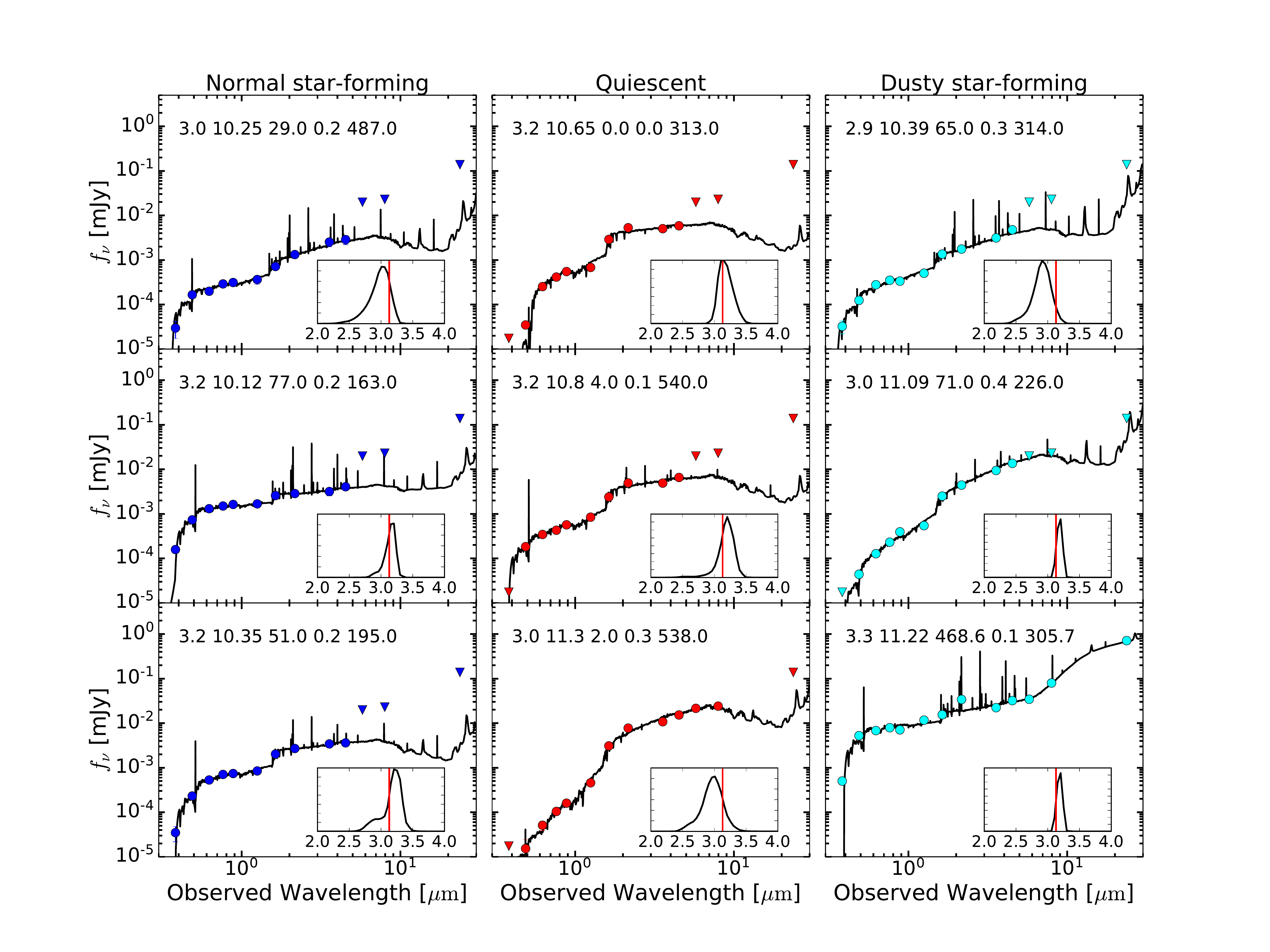}
\caption{
CIGALE SED fitting results for a sub-sample of our photo-$z$ galaxies, which include normal star-forming galaxies (left), quiescent galaxies (middle) and dusty star-forming galaxies (right). The black solid lines are the best-fit model spectra. Filled circles represent the observed fluxes, while triangles denote 2$\sigma$ upper flux limits in the case of nondetection. In the inset of each panel, we also show the probability distribution function of the photometric redshift for each galaxy, and the redshift of the protocluster is shown as a red vertical line. On the top of each subpanel, we list the best-fit photo-$z$, log($M_{star}$) (in units of $M_\sun$), SFR (in units of $M_\sun$ yr$^{-1}$), dust reddening parameter E(B-V) and age (in units of Myr).
}
\label{fig:sed}
\end{figure*}

We investigate the physical properties of different galaxy populations classified above. 
Figure \ref{fig:sed} shows the SED fitting results for a sub-sample of our photo-$z$ candidates. We see that quiescent galaxies are distinguished by their prominant break between $J$ and $K_S$, while dusty star-forming galaxies usually are redder beyond NIR wavelength as indicated by their best-fit spectra. 

%It can be seen that the normal star-forming galaxy candidates are usually less massive ($<10^{10.5}$ $M_\sun$) with younger stellar populations, and are forming stars in a moderate rate (several tens of solar mass per year) with moderate amount of dust. On the other hand, the quiescent galaxy candidates in our sample are usually more massive galaxies ($>10^{10.5}$ $M_\sun$) with older stellar populations that have prominant Balmer/4000 $\mathrm{\AA}$ break between $J$ and $K_S$. They have mostly stopped their star-formation as indicated by their very low SFRs. Finally, dusty star-forming galaxies selected by our criteria tend to be massive and forming stars at much higher rate (up to several hundreds of solar mass per year) than the normal ones, which are usually accompanied with a lot of dust.

\begin{figure*}[ht!]
\epsscale{1.2}
\plotone{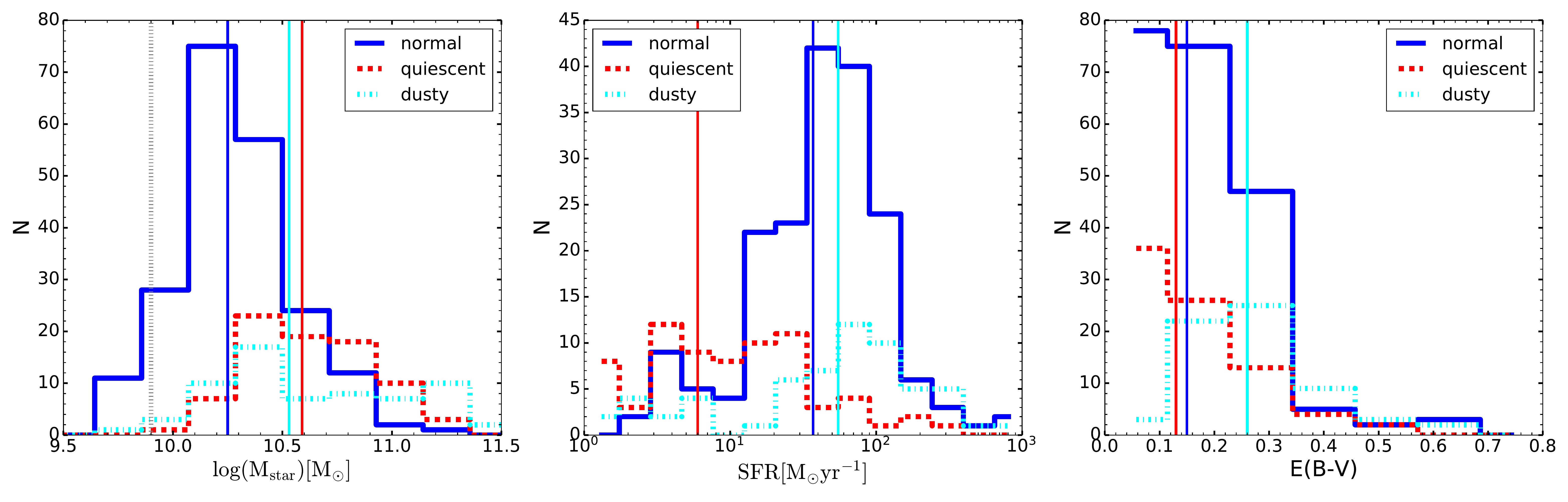}
\caption{
 {\it Left:} The stellar mass distributions of different galaxy populations. The blue, red, cyan lines corresponds to normal star-forming galaxies, quiescent galaxies and dusty star-forming galaxies respectively. The grey vertical line is the stellar mass completeness limit for our photo-$z$ sample. The solid vertical lines represent the median value for corresponding galaxy population.
 {\it Middle:} The distributions of SFRs of different galaxy populations.
 {\it Right:} The distributions of dust attenuation E(B-V) of different galaxy populations.
}
\label{fig:hist_div}
\end{figure*}

Figure \ref{fig:hist_div} shows the stellar mass, SFR and dust reddening E(B-V) distributions of different galaxy populations selected by our criteria. It can be seen that the stellar masses of quiescent and dusty galaxies are skewed towards higher mass end: the median stellar masses of the quiescent and dusty galaxies are 10$^{10.59}$ $M_\sun$ and 10$^{10.53}$ $M_\sun$, respectively, while only 10$^{10.25}$ $M_\sun$ for normal star-forming galaxies. The quiescent galaxy sample has a very low median SFR of 6 $M_\sun$ yr$^{-1}$, comparing to the normal star-forming galaxy population which has a median SFR of 37 $M_\sun$ yr$^{-1}$. The dusty star-forming galaxies are skewed towards higher SFR end, with a median value of 55 $M_\sun$ yr$^{-1}$. As for the dust content, dusty star-forming galaxies have a higher dust extinction with a median E(B-V) $=0.26$, while quiescent and normal star-forming galaxies are less obscured by dust with a median E(B-V) of $0.13$ and $0.15$ respectively. We also validate the difference of the three galaxy populations using the K-sample Anderson-Darling test \citep{ADtest}. This test is similar to the commonly used Kolmogorov-Smirnov test but more sensitive and can deal with more than two samples. The Anderson-Darling test finds significant differences among the three populations: $p$-value$<$0.001 in all cases for the mass, SFR and E(B-V) distributions.

\begin{figure*}[ht!]
\epsscale{0.7}
\plotone{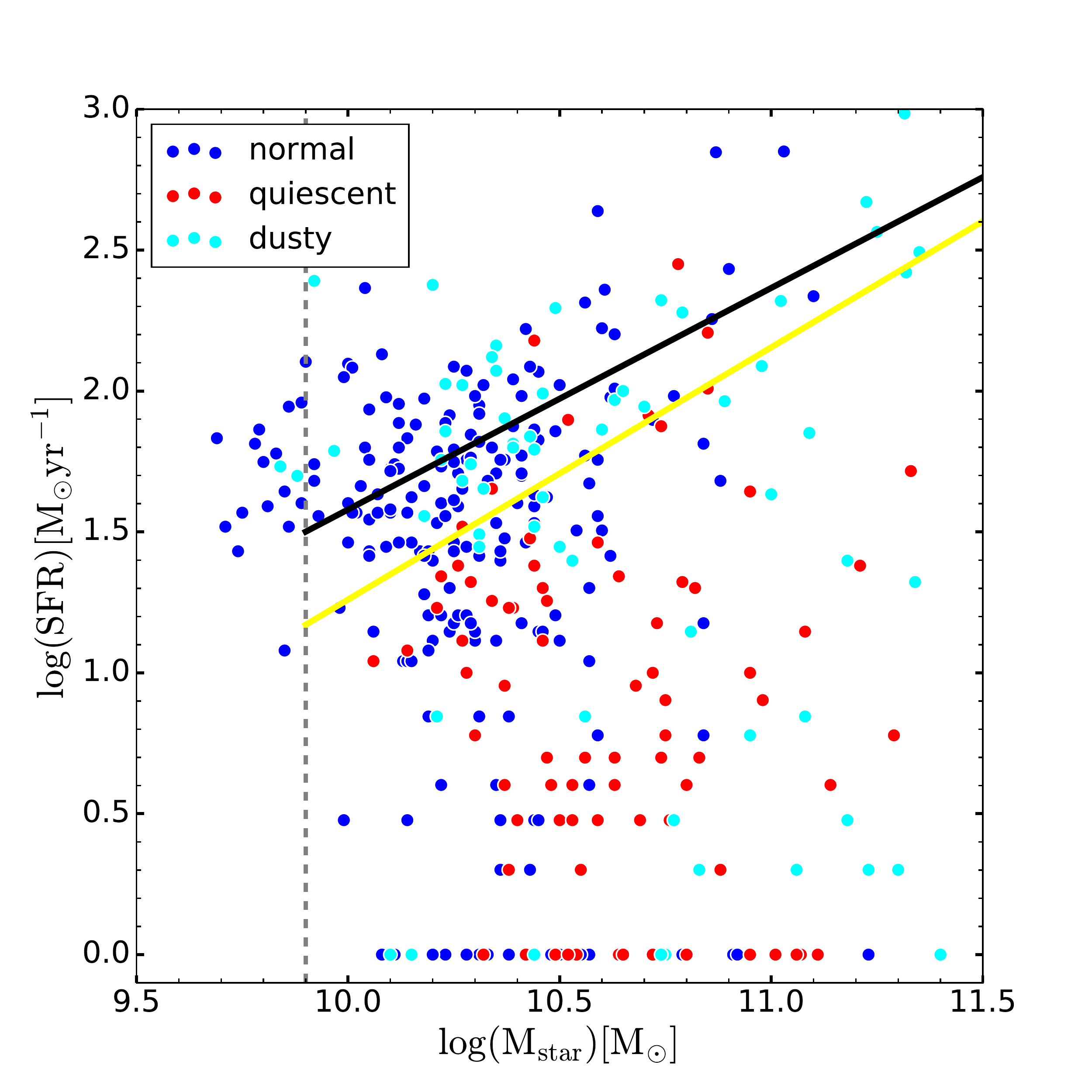}
\caption{SFR-$M_\mathrm{star}$ relation for different galaxy populations in our sample. The black solid line is the observed relation based on the calibration of \cite{Speagle14} at $z=3.1$, while the yellow solid line represents that from a semi-analytic model by \cite{Dutton10} at $z\sim3$. The vertical dashed line is the mass completeness limit of our sample. Galaxies with SFR~$=0$ are indicated in the log(SFR)~$=0$ location.
}
\label{fig:msd}
\end{figure*}

Numerous studies have indicated a correlation between  stellar mass ($\mathrm{M_{star}}$) and SFR for star-forming galaxies, which is the so-called star-forming main sequence (MS) \cite[e.g.,][]{Noeske07,Elbaz07,Daddi07,Rodighiero11,Reddy12,Speagle14,Salmon15,Santini17}.
In Figure \ref{fig:msd}, we show the locations of our photo-$z$ galaxies on the SFR-$\mathrm{M_{star}}$ plane.  In the figure we also show the MS relation from both observation and simulation at $z\sim3$. On one hand, most of our star-forming galaxies and dusty star-forming galaxies are located close to the MS from both observation \citep{Speagle14} and simulation \citep{Dutton10}. On the other hand, the majority of the quiescent galaxy candidates lie below the MS relation: among the 81 candidates, only 7 lie above \cite{Dutton10} relation while only 2 lie above \cite{Speagle14} relation. Therefore, our quiescent galaxy candidates are indeed quenched systems with little on-going star-formation activities, compared to the star-forming galaxy populations. This further justifies our selection criteria for different galaxy populations.

\section{sky distribution of galaxies}\label{distribution}
In \cite{Shi192}, we compared the sky distributions of both LBGs and LAEs and found a spatial offset between the overdensities traced by these two different galaxy populations, which may indicate different halo formation time or certain environmental effects.
To further investigate this problem, in the section we discuss the spatial distribution of the photo-$z$ galaxies in our field. 

\begin{figure*}[ht!] 
\epsscale{1.0}
\plotone{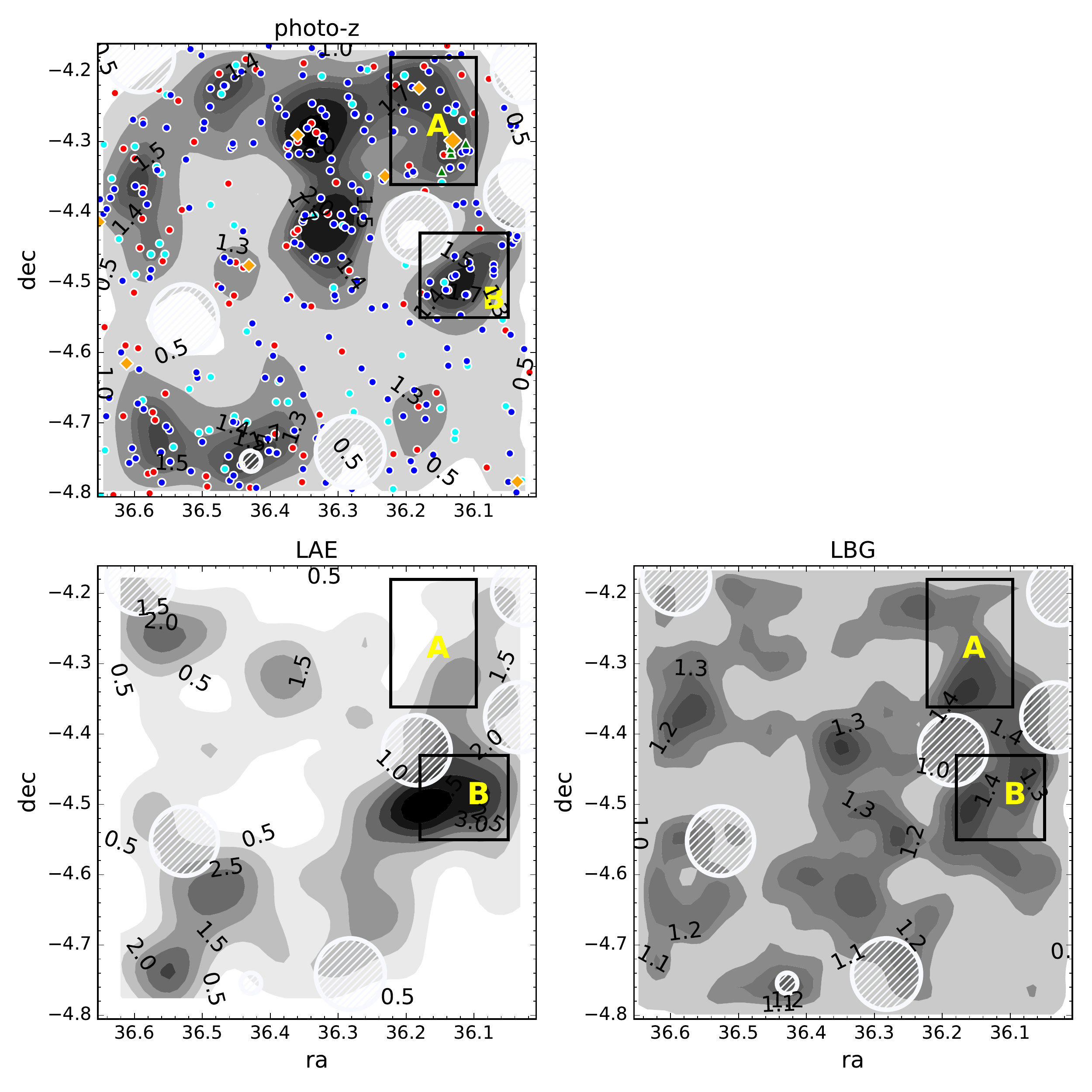}
\caption{{\it Top left:} sky distributions of the photo-$z$ galaxies ($2.9<z<3.2$). The blue, red and cyan circles indicate the normal star-forming galaxies, quiescent galaxies and dusty star-forming galaxies respectively. The orange diamonds are the sources that detected in 24$\micron$ (the largest one denotes the BCG candidate) while the green triangles are the spectroscopically confirmed LBGs in \cite{Toshikawa16}. The contours are constructed by smoothing the photo-$z$ galaxies with a Gaussian kernel of FWHM=10 Mpc. The white hatched circles are the masked area near bright saturated stars. The two boxes represent the overdense protocluster regions (20 Mpc in length each) and are dubbed as `A' and `B'. {\it Bottom left:} LAE density map smoothed using a FWHM=10 Mpc Gaussian kernel. {\it Bottom right:} LBG density map smoothed using a FWHM=6 Mpc Gaussian kernel.
}
\label{fig:density}
\end{figure*}

In the top panel of Figure \ref{fig:density}, we show the sky distributions of the photo-$z$ galaxies. The surface density map is created using a Gaussian smooth kernel of a FWHM of 10 Mpc which is the same as the smoothing scale used in \cite{Shi192} for the LAE density map. We choose this smoothing scale as the number of photo-$z$ galaxies are comparable with the LAEs while much less than the number of LBGs ($\sim7000$) for which a 6 Mpc smoothing scale has been used \citep{Shi192}. The contour line values represent the local surface density relative to the field average. Different galaxy populations are also indicated in different colors. For comparison, in the bottom two panels, we reproduce the density maps of LAEs and LBGs used in \cite{Shi192}.

For the photo-$z$ galaxies, there appear to be several large overdensities in the field: two in the middle of the north and two in the west. The overdensity in the mid-west of the field roughly coincides with the known LAE overdensity and southern LBG overdensity, while the overdensity area in the northwest is  largely co-spatial with the northern LBG overdensity. To make a uniform selection, we choose the area within the 1.3$\Sigma$ iso-density contour line as the overdense regions in the mid-west that cohabit with the LBG and LAE overdensities. In Figure \ref{fig:density} we mark the two overdensities using box `A' and `B' that enclose the  1.3$\Sigma$ iso-density contour lines. The left boundary of the box `A' is used to cut the 1.3$\Sigma$ iso-density contour line to make a closed region. The center of the box `A' and `B' is at [36.15875, -4.27073] and [36.11415, -4.48938] in R.A. and decl., respectively. 

There are also two overdensities in the middle of the field. In particular, there appears to be a similar peak in the middle of the LBG map (bottom right panel of Figure \ref{fig:density}). We speculate several possibilities for the overdensities in the middle. One is that these overdensities are simply coincidental alignment of galaxies along the line of sight which have no physical associations. Since the LBGs range from $z=2.8\sim3.5$ \citep{Shi192} which are far larger than the typical protocluster size of 20 Mpc ($\Delta z\sim0.02$) at $z\sim3$ \citep{Chiang13}, this possibility is non-trivial. The same is true for our photo-$z$ galaxies ($\Delta z=0.4$). Furthermore, as our photo-$z$ galaxies are selected in a different way than the LBGs (they are more massive, NIR luminous galaxies), this could result in different distribution of galaxy populations seen in the surface density map, such as the mid-north peak which is present in the photo-$z$ map but absent in the LBG map. Another possibility is that these are genuine (proto)clusters at different redshift than $z=3.13$. The LAE map targets at the $z=3.13\pm0.02$ structures \citep{Shi192} in which we do not find any significant overdensities in the middle of the field (bottom left panel of Figure \ref{fig:density}). There is a small peak in the mid-north which is a bit offset from the photo-$z$ overdensity but too weak to be considered as a real structure in comparison to the major one in the west, even to other small peaks in the map. Therefore, these photo-$z$ peaks could be other structures located in $z\sim2.9-3.3$ but at different redshift than the LAE and spectroscopically confirmed LBGs at $z=3.13$. At current stage, in lack of spectroscopic observations in the mid-north of the field, we leave it to future studies and only regard the ones in the west (`A' and `B') as potential protocluster regions at $z=3.13$ in the remainder of this paper.

There are 31 photo-$z$ galaxies within the 1.3$\Sigma$ iso-density contour in `A', among which 18 are normal star-forming galaxies, 5 are quiescent galaxy candidates and 8 belong to the dusty star-forming galaxy population. Thus nearly half of the the galaxies (42$\pm$12\%, where the error denotes the Poisson noise) are evolved and/or dusty galaxy candidates. On the other hand, 18 galaxies reside within the 1.3$\Sigma$ iso-density contour in the `B' region, with 2 being quiescent galaxies, 2 being dusty star-forming galaxies and 14 being normal star-forming galaxies: evolved and/or dusty galaxies only take up 22$\pm$11\% of the total in this case. It appears that region `A' is dominated by more evolved and/or dusty galaxy populations while region `B' mainly contains normal star-forming galaxies. It is noteworthy that region `B' largely coincides with the major LAE overdensity in the bottom left panel of Figure~\ref{fig:density}, supporting the general idea that LAEs are young star-forming galaxies with little dust obscuration.

We use an angular two-point cross-correlation function (CCF) to quantify the spatial correlation between the LAEs and different photo-$z$ galaxy populations. The angular two-point correlation function is often used to describe the excess probability of finding two galaxies separated by a certain angular distance, relative to the random distributions, which has been used in the literature to investigate the spatial cohabitation between different galaxy populations \citep[e.g.,][]{Tamura09,Harikane19}. We calculate the CCF using the \cite{Landy93} estimator:
\begin{equation*}
\omega(\theta)=\frac{D_1D_2(\theta)-D_1R_2(\theta)-R_1D_2(\theta)+R_1R_2(\theta)}{R_1R_2(\theta)},
\end{equation*}
where $DD$, $DR$, $RD$, $RR$ are the galaxy-galaxy, galaxy-random, random-galaxy and random-random pair counts respectively, for group 1 and 2. The statistical errors of the CCFs are estimated from the standard deviation of 1,000 bootstrap realizations.

\begin{figure*}[ht!]
\epsscale{1.0}
\plotone{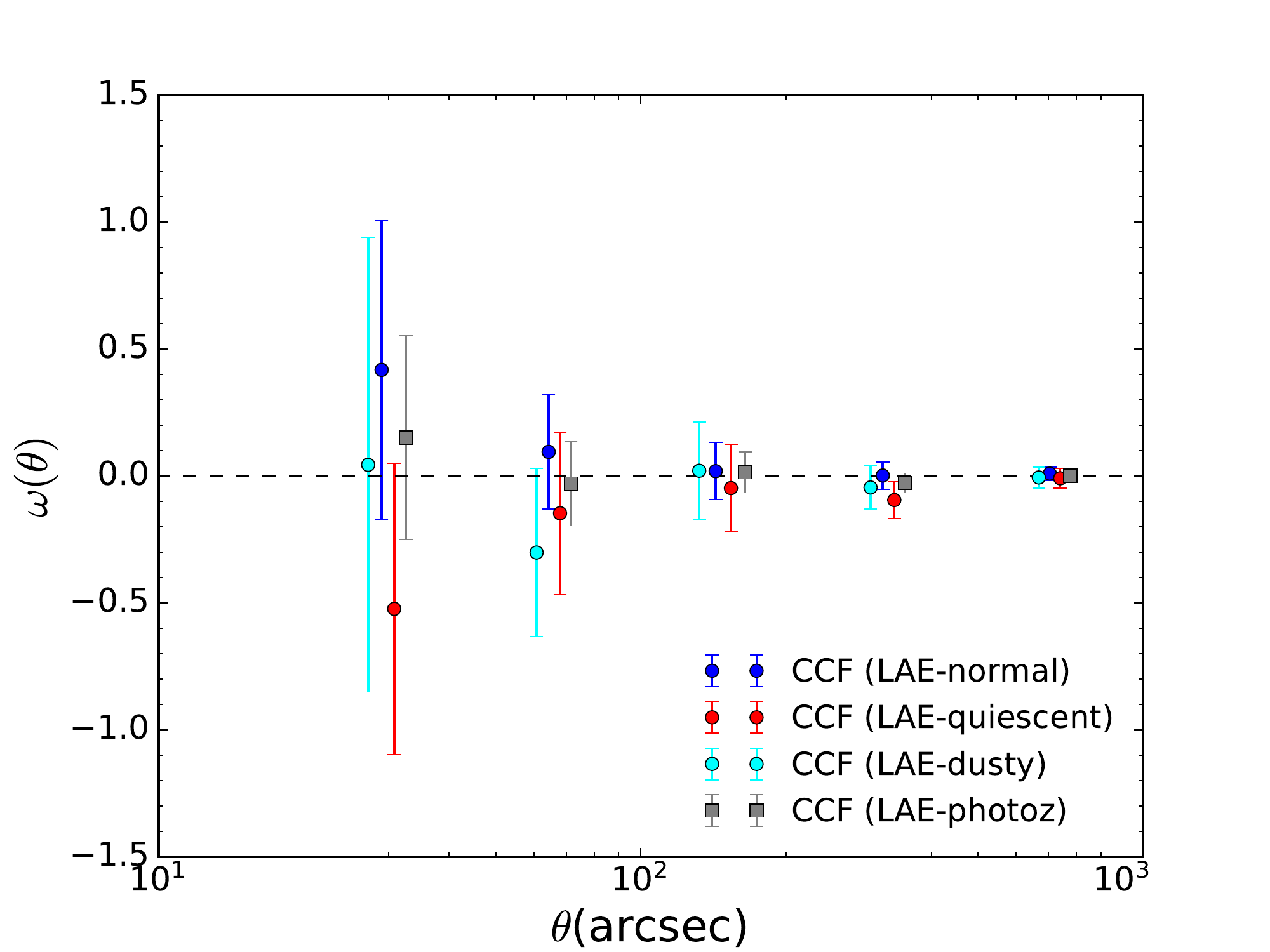}
\caption{Angular cross-correlation between different galaxy populations with the LAEs. The blue, red, cyan circles show the CCFs between normal star-forming galaxies and LAEs, quiescent galaxies and LAEs, dusty star-forming galaxies and LAEs respectively. The grey points show the CCFs between all the photo-$z$ galaxies and LAEs. The errors are estimated from the standard deviation of 1,000 bootstrap samples.
}
\label{fig:corrfunc}
\end{figure*}

Figure \ref{fig:corrfunc} shows the CCFs between the LAEs and different photo-$z$ galaxy populations in the entire field. We find a strong correlation between the LAEs and the normal star-forming galaxies at small angular scales, suggesting close association of these two populations. Meanwhile, there is no obvious correlation between the LAEs and the dusty star-forming galaxy candidates. There also appears to be an anti-correlation between the LAEs and the quiescent galaxy population. This agrees with our visual impression.

All in all, our results suggest the presence of two photo-$z$ galaxy overdensities, which are co-spatial with previously identified LAE and LBG overdensity in the field. One galaxy overdensity is dominated by normal star-forming galaxies while the other contains a large fraction of quiescent/dusty galaxies. We discuss possible implications of our results in Section 5.2.

\section{Discussion} \label{dis}
\subsection{Environmental Impact on Galaxy Properties} \label{env}
Having identified the high-density protocluster regions in Section \ref{distribution}, we investigate the impacts of local environment of the protocluster on the physical properties of galaxies in this section.

The main challenge in studying the environmental effects on protocluster galaxies is the lack of spectroscopic redshifts. The large redshift dispersion of our photo-$z$ galaxies ($\Delta z \sim0.4$) prohibits the precise determination of the galaxy membership and a robust mapping of the genuine protocluster region. Thus follow-up spectroscopic observations on this protocluster are urgently needed.

With the above caveats in mind, we compare the physical properties of protocluster galaxy candidates with those in the field. To this end, we divide our photo-$z$ sample into two subsamples: the `overdensity' sample within the 1.3$\Sigma$ iso-density contour lines enclosed by the box `A' and `B' shown in Figure \ref{fig:density} and the `field' sample that is simply all the 356 photo-$z$ galaxies in the entire field.  As discussed in Section \ref{distribution}, since this paper focuses on the $z=3.13$ confirmed protoclusters in `A' and `B', also in lack of spectroscopic information elsewhere, we do not consider other apparent `overdense' regions. Because we cannot rule out the possibility that other `overdense' regions are genuine structures at other redshift, we choose the whole survey field (including `A' and `B') as the general field.  This definition of the field should represent the average galaxy pupulations at $z\sim3$. The `overdensity' sample is further devided into two groups (region `A' and `B' respectively). Our photo-$z$ galaxies are all selected from the same set of photometric data and their properties are determined using the same method, therefore no selection effect is needed to be accounted for. 

Fourty-nine galaxies are located within the high-density regions (31 in region `A' and 18 in region `B'). Table \ref{table2} lists the median physical properties obtained of each subsample. The errors correspond to the median absolute deviations which are less affected by outliers.

\begin{figure*}[ht!]
\epsscale{1.0}
\plotone{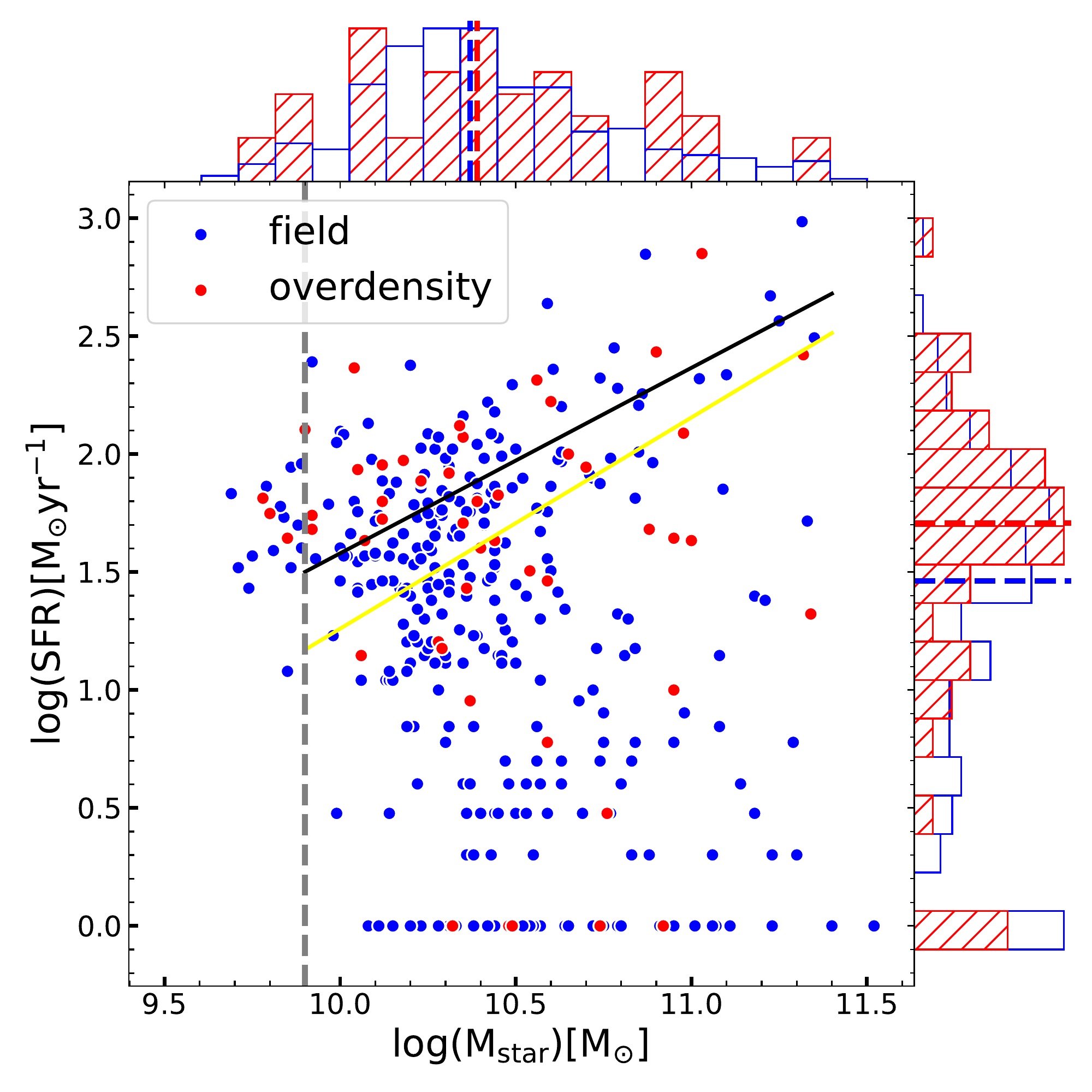}
\caption{SFR-$M_\mathrm{star}$ relation for the protocluster and field galaxies in our sample. The black solid line is the observed relation based on the calibration of \cite{Speagle14}, while the yellow solid line represents that from a semi-analytic model by \cite{Dutton10}. The vertical dashed line is the mass completeness limit of our sample. Galaxies with SFR~$=0$ are indicated in the log(SFR)~$=0$ location. The normalized histograms show the distributions of SFR and stellar mass of the two groups, with the vertical lines indicating the median values.
}
\label{fig:ms}
\end{figure*}

\begin{figure*}[ht!]
\epsscale{1.2}
\plotone{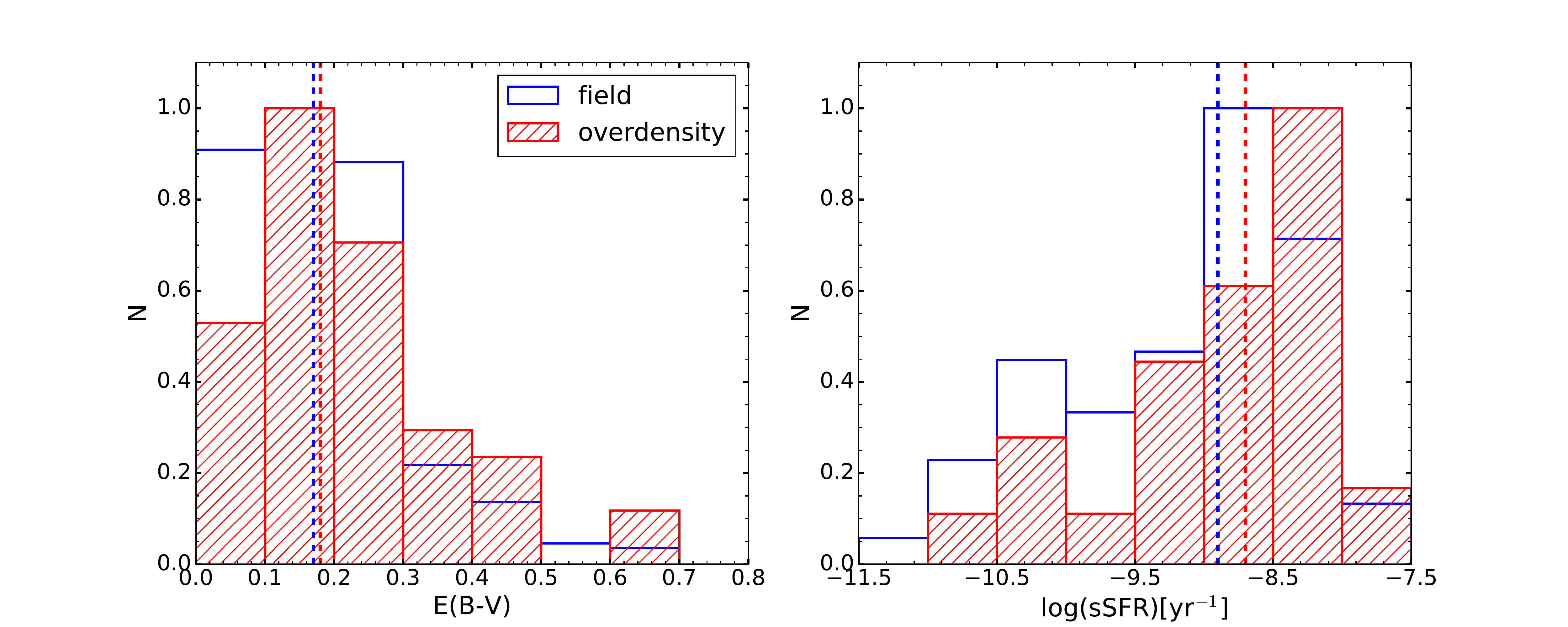}
\caption{Normalized histograms of dust extinction E(B-V) (left) and sSFR (right) for the protocluster and field galaxies. The vertical dashed lines represent the median values of each group. 
}
\label{fig:hist_env}
\end{figure*}

In terms of stellar mass, we do not find obvious differences between different subsamples. A two-sample Kolmogorov--Smirnov (K-S) test cannot distinguish between region `A' and/or `B' with the field, as well as between `A' and `B' ($p$-value $>$ 0.8 in all cases), which is also confirmed in Table \ref{table2}. As for the star-formation rate, on one hand, there is no significant difference between `A' and the field ($p$=0.2), while the K-S test indicates there is a significant difference between `B' and the field ($p$=0.003). However, when we compare only `A' with `B', the difference fades away, with a $p$-value of 0.3. If `A+B' is compared with the field, K-S test suggests the probability that they come from the same underlying distribution is $<1\%$ ($p=0.004$). 
In Figure \ref{fig:ms}, the photo-$z$ protocluster galaxies (`A+B') and field galaxies are shown on the SFR-M$_\mathrm{star}$ plane. It can be seen that although the stellar masses of the two groups are similar in distribution, the star-formation rates of the protocluster galaxies are skewed towards higher values than the field counterparts. The enhancement of the SFRs can be further seen in the right panel of Figure \ref{fig:hist_env}, where we show the distributions of the specific star-formation rate (sSFR, defined as SFR/M$_\mathrm{star}$) for the two groups. The K-S test implies strong distinction between `B' and the field ($p$=0.02) while no significant difference between `A' and the field is observed ($p$=0.6). This leads to a moderate difference between the overall overdensity with the field ($p$=0.07), but the K-S test cannot reject the null hypothesis that `A' and `B' come from the same distribution ($p$=0.2). These K-S tests appear to suggest that galaxies within the protocluster regions are forming stars more actively than the general field. In Table \ref{table2}, we can also see that SFR in the protocluster regions `A+B' is enhanced by $\sim76\%$ as compared to the field. This elevation is even higher for `B' ($\sim124\%$) than that for `A' ($\sim52\%$). 

As for the dust extinction, no significant difference between `A' and/or `B' with the field is observed ($p>0.2$). Figure \ref{fig:hist_env} (left panel) shows the distributions of dust attenuation parameter E(B-V) between the protocluster galaxies and field galaxies. Both Table \ref{table2} and the histogram imply that the overdense protocluster regions have the similar dust content as the general field.

The enhancement of SFRs in the protocluster regions is consistent with our previous work \citep{Shi192} where we found that the Ly$\alpha$ luminosity and UV luminosity of the protocluster galaxies have higher median values than the field. In \cite{Shi192}, since we did not have mass measurement on the LAEs, we could not rule out the possibility that the trend is due to a deficit of low-mass galaxies in the protocluster environment. This work takes a step further by measuring the stellar masses, confirming the enhancement of SFRs of the protocluster galaxies are not due to the lack of low-mass galaxies but an overall boost of star-formation efficiency in the overdense protocluster environment. Furthermore, our results appear to suggest that protocluster galaxies in region `B' have even higher star formation efficiency, compared to those in region `A'. We will discuss this later in Section 5.2.

Recently, \cite{Shimakawa18} studied a protocluster at $z=2.5$ using H$\alpha$ emitters (HAEs) to trace the large scale structure. They found that HAEs in the densest regions of the protocluster have enhanced SFRs and dust extinctions at high confidence level, indicating a rapid mass assembly of star-forming galaxies in the protocluster regions. At the similar redshift, \cite{Wang18} investigated the molecular gas properties of a distant X-ray cluster \citep{Wang16}, finding that the star-formation efficiency (indicated by the ratio between SFR and gas mass) is elevated in the cluster region in comparison with the field. They argued that the galaxies in the central regions of this cluster will consume all the gas and become quiescent in a short time scale. The enhancement of star-formation activity in our protocluster field (both measured from LAEs and photo-$z$ galaxies) is consistent with the above studies. We argue that galaxies in our protocluster regions are also experiencing an accelerated mass assembly, likely  consuming their gas rapidly and becoming quiescent in a short time period.

It is known that protoclusters often host extremely dusty star-forming galaxies such as submillimeter galaxies (SMGs) with SFRs exceeding 1000 solar masses per year \citep[e.g.,][]{Kato16,Casey16,Miller18,Oteo18,Cheng19}. These objects are very luminous in submillimeter wavelength and are heavily dust obscured which are generally invisible in rest-frame UV-NIR wavelengths. It is possible that some SMGs exist in our protocluster that are missed by our selection. An extensive study of this galaxy population requires future submillimeter observations in this field. If these dusty starbursts systems are confirmed to be preferably concentrated in our protocluster regions, the enhancement of star-formation activities in the dense environments would be even higher.

\subsection{Difference in Galaxy Constituents of the Two Overdensities}
In Section \ref{distribution}, we find that the photo-$z$ galaxies form two overdensities `A' and `B' in the field, which are co-spatial with our previously identified LBG and LAE overdensities respectively \citep{Shi192}. Given the large end-to-end size of these two overdensities ($\sim$40 Mpc) that is almost twice the size of the largest protocluster in \cite{Chiang13} at $z\sim3$ ($\sim$20 Mpc), we assume these two overdensities trace separate structures in the following discussion.

In \cite{Shi192}, we argued that the spatial segregation of `A' and `B' is possibly due to different formation time of underlying dark matter halos. The former structure formed earlier than the latter, thus is traced by older, more massive LBGs while the latter traced by younger LAE population. In this work, our galaxy selection criteria suggest that 42$\pm$12\% of the galaxies in `A' are massive quiescent and/or dusty galaxies (similar to the field of 41$\pm$3\%), comparing to 22$\pm$11\% in `B' (Section  \ref{distribution}).  Meanwhile, the normal star-forming galaxies are strongly correlated with the LAEs as shown in Figure \ref{fig:corrfunc}. Therefore, it turns out that the overdensity `B' largely coincides with the LAE overdensity (see Figure \ref{fig:density}). 
 It is also noticed that there are two 24$\micron$ detected objects (orange diamonds in Figure \ref{fig:density}) located in `A' including one brighest cluster galaxy candidate discovered in \cite{Shi192}, while no source is detected at 24$\micron$ in  `B'. In addition, in Section 5.1 we see that galaxies in `B' appear to have higher star-formation efficiency than those in `A'. Taken together, these results suggest that the region `A' is a more evolved structure which is mainly traced by old and/or dusty galaxy populations.  In comparison, region `B' formed at a later stage and is dominated by younger galaxy population which consists mostly of normal star-forming galaxies such as LAEs. The elevated star-formation activities in `B' suggests that its galaxy constituents are rapidly building their masses. In comparison, `A' appears to be a more settled structure that already passed the peak of its star-formation.

Last but not least, we consider the scenerio that `A' and `B' are two protoclusters that embedded in a primodial supercluster.  In the local and nearby universe, a supercluster typically consists of a group of galaxy clusters, which forms the largest structures residing in the filaments of the cosmos \citep[e.g.,][]{Abell58,Chon13,Tully14}. Although the definition of superclusters is not precise, the size of a supercluster can range from several tens Mpc to more than one hundred Mpc \citep{Chon13}. At high redshift ($z>2$), several potential primodial superclusters have been reported \citep[e.g.,][]{Ouchi05,Dey16,Topping16,Cucciati18,Toshikawa19}. Especially, in the same CFHTLS D1 field, \cite{Toshikawa19} recently found evidence for presence of a primeval supercluster at $z\sim4.9$ within a volume of $\sim33\times12\times64$ Mpc$^3$. Based on follow-up spectroscopic observations of the LBGs selected in \cite{Toshikawa16}, they identified three overdense structures with a redshift separation $\Delta z\sim0.05$ between each density peak. They argued that these structures will evolve independently and become part of a supercluster by $z=0$. These studies suggested that premordial superclusters appear at high redshift in parallel with the formation of its cluster/group components.

In this work, the end-to-end distance between the two structures is $\sim$40 Mpc, comparable to that of the nearby superclusters while too large for a typical protocluster ($\sim20$ Mpc). The total mass of the two structures is $\sim10^{15}\mathrm{M_\sun}$ each, as estimated in \cite{Shi192}. The mass of the LAE structure was calculated using the observed galaxy overdensity and its enclosed volume, while for the mass of the LBG structure we used simulation to infer its intrinsic overdensity along with the assumed volume to get an estimate.  The `A' region has five spectroscopically confirmed LBGs at $z=3.13$ identified in \cite{Toshikawa16} (green triangles in Figure \ref{fig:density}), while `B' is dominated by a large population of LAEs at $z=3.132\pm0.023$. Considering their similar mass and redshift but large transverse separation, it is likely that `A' and `B' will grow independently into two separate massive clusters as part of a supercluster by $z=0$. Only future spectroscopy in these regions can elucidate the true underlying large-scale structure.

\begin{deluxetable}{cccccc}[h]
\tablecaption{Physical properties of the photo-$z$ galaxies in different environments \label{table2}}
\tablehead{
\colhead{Region} & \colhead{N} & \colhead{log(Mass)} & \colhead{SFR} & \colhead{E(B-V)} & \colhead{log(sSFR)} \\
\colhead{} & \colhead{} & \colhead{($\mathrm{M_\sun}$)} & \colhead{($\mathrm{M_\sun yr^{-1}}$)} & \colhead{} & \colhead{($\mathrm{yr^{-1}}$)}
}
\startdata
A & 31 & 10.39$\pm$0.46 & 44$\pm$44 & 0.17$\pm$0.13 & -8.9$\pm$0.7 \\
B & 18 & 10.38$\pm$0.33 & 65$\pm$35 & 0.22$\pm$0.12 & -8.4$\pm$0.4 \\
A$+$B & 49 &10.39$\pm$0.40 & 51$\pm$53 & 0.18$\pm$0.12 & -8.7$\pm$0.7 \\
field & 356 & 10.37$\pm$0.30 & 29$\pm$39 & 0.17$\pm$0.13 & -8.9$\pm$0.8
\enddata

\end{deluxetable}

\subsection{Search for Rare Sources in the Protocluster Field}
Above we have selected the protocluster galaxy candidates using the available optical-IR (OIR) data. Apart from these OIR sources, dense protocluster environments are often found to host powerful radio galaxies or X-ray luminous AGNs \citep[e.g.,][]{Venemans05,Overzier06,Miley08,Hayashi12,Cooke14,Digby10,Kubo13,Krishnan17}, thus it is interesting to search for these rare sources in our protocluster to look for a sign of enhanced AGN activities.

First, we cross-match our photo-$z$ sources with the new XMM-Newton point-source
catalog from from the XMM-SERVS survey \citep{Chen18}. Their catalog has 5,242 sources detected in the soft (0.5--2 keV), hard (2--10 keV),
and full (0.5--10 keV) bands, which reaches a flux limit of 1.7$\times10^{-15}$, 1.3$\times10^{-14}$, and 6.5$\times10^{-15}$ erg cm$^{-2}$ s$^{-1}$ respectively. Using a matching radius of 1.5$\arcsec$, no counterpart in our photo-$z$ sample is found. It is possible that some faint X-ray sources in our sample are simply missed by their detection, as the X-ray sources in the famous SSA22 protocluster from the {\it{Chandra}} catalog \citep{Lehmer09} have an average value of 1.6$\times10^{-15}$ erg cm$^{-2}$ s$^{-1}$ for the full band, which lie well below the sensitivity of the XMM-SERVS survey. In order to further investigate the potential X-ray signals in our protocluster field, future deep X-ray surveys are needed.

Second, we also search for radio counterparts in our photo-$z$ sample, using the publicly available radio catalog obtained from the Very Large Array (VLA) at 1.4 GHz covering our field \citep{Bondi03}. The catalog contains radio sources down to a 5$\sigma$ depth of $\sim$0.08 mJy. We find four counterparts in our photo-$z$ sample within a 1.5$\arcsec$ search radius whose total flux densities are in the range of 0.09--3.30 mJy. However, none of these sources is in the overdense protocluster regions (`A' or `B'). Therefore our protoclusters may generally lack of luminous ($\gtrsim0.1$ mJy) radio sources. 

%3.30 mJy, 0.09 mJy, 0.21 mJy and 0.24

Last, we further investigate the brightest cluster galaxy candidate (BCG) found in our previous study. In \cite{Shi192}, we discovered an ultra massive galaxy G411155 which lies very close to the spectroscopic sources in the `A' region (see Figure~\ref{fig:density}). G411155 is the brightest source in our LBG catalog and also the reddest. Our preliminary SED-fitting result using the \cite{Bielby12} and \cite{Lonsdale03} catalogs suggested this galaxy is dominated by a dust obscured AGN and is in a phase of intense star-formation.

In this work, using our improved PSF-matched photometry from optical to IR, we re-visit the physical properties of G411155. Our photometry shows that this galaxy has a $K_S$ magnitude of 21.07, consistent with that of \cite{Bielby12} catalog. We obtain $J-K_S=2.15$ for this galaxy from our photometry which is larger than that of \cite{Bielby12} ($J-K_S=1.92$). This extremely red color places G411155 further into the category of hyper extremely red objects (HEROs) ($J-K_S>2.1$) \citep{Totani01}, which are thought to be  primordial elliptical galaxies that still in the phase of dusty starburst. Our updated SED-fitting on this source yields a stellar mass of $1.0\times10^{11}~\mathrm{M_\sun}$ with  SFR of $\sim123~\mathrm{M_\sun}$ yr$^{-1}$. The age of G411155 is $\sim500$ Myr which is older than previous estimate ($\sim$200 Myr). SED-fitting also suggests that 80\% of its IR luminosity is dominated by a dust obscured AGN. In the entire field of this work (1,156 arcmin$^2$), G411155 is the only object that meets the HERO selection criterion with mass $\geq 10^{11} \mathrm{M_\sun}$ and SFR $>100~\mathrm{M_\sun}$ yr$^{-1}$. We conclude that G411155 is a rare source in the protocluster region and even in the field. It is likely that we are witnessing the formation of a BCG in the protocluster region `A'. Future follow-up spectroscopic observations using telescopes such as Keck/MOSFIRE or JWST are needed to confirm this BCG and  provide us with more information of its properties.

\section{Summary}\label{sum}
In this work, by utilizing the multiwavelength data in the CFHTLS D1 field around a protocluster `D1UD01', we identify 3.6 $\micron$-selected galaxies that are candidate members of the protocluster with the help of photometric redshift.  We divide them into different categories and study their physical properties, trying to understand the spatial configuration of the underlying large-scale structure in and around the protocluster, and to further investigate the environmental impact on galaxy formation. Based on our analysis, we conclude the following:

1. Diverse galaxy populations have been found in the protocluster field, including normal star-forming galaxies, massive quiescent galaxies and dusty star-forming galaxies. With only 33\% of the photo-$z$ galaxies satisfying the LBG criteria, our sample includes a high abundunce of massive galaxies ($\gtrsim10^{10}~\mathrm{M_\sun}$)  that are generally missed from previous rest-frame UV-selected star-forming galaxies such as LBGs and LAEs. The LAEs in \cite{Shi192} appear to be spatially correlated with the normal star-forming galaxies in our sample, but not with the more massive quiescent and/or dusty star-forming galaxies, suggesting that LAEs are biased tracers of the underlying large-scale structure which typically miss the more massive quiescent and/or dusty galaxy populations that are likely to be present in protoclusters.

2. We identify two significant photo-$z$ overdensities around the protocluster region. The northern overdensity `A' is largely co-spatial with the largest LBG overdensity, and consists of a high fraction (42\%) of quiescent and/or dusty galaxies. The southern structure `B' overlaps with the LAE overdensity and contains a much lower fraction (22\%) of quiescent and/or dusty galaxies. Our result is consistent with \cite{Shi192}, where we argued that traced by older and more massive galaxy populations, `A' is a more evolved structure than `B'.  Given the large size and transverse separation of the two structures, it is likely that `A' and `B' may represent two distinct protoclusters that are in different formation stages, which will evolve into a supercluster by present day.

3. We find strong evidence that the average star-formation activities are enhanced in the protocluster regions in comparison with the field. Although having similar masses, the protocluster galaxy members have higher SFRs than the field galaxies by $\sim$76\%, which confirms our previous study based on LAEs \citep{Shi192}. We argue that the protocluster galaxies are in a phase of accelerated mass assembly, rapidly consuming their gas content and will likely become quiescent in a short time period.

4. We do not find any X-ray or radio luminous sources in our photo-$z$ sample. However the absence of these rare sources could be due to the low sensitivity of the current available observations, which calls for future deep surveys in this field. We also confirm that the brightest cluster galaxy candidate discovered in our previous study is indeed a rare and unique source in the protocluster field. Further spectroscopic validation of this galaxy is still needed to determine whether it truely belongs to the protocluster. 

We thank the anonymous referee for a careful reading of the manuscript and insightful comments. We are also grateful to the statistic editor for introducing new statistical methods in our analysis. This work is supported by the National Key R\&D Program of China No.\:2017YFA0402600, and NSFC grants No.\:11525312, 11890692.

\bibliography{xmu}
\bibliographystyle{aasjournal}



\end{document}